\documentclass[sigplan,twocolumn]{acmart}
\renewcommand\footnotetextcopyrightpermission[1]{}
\usepackage{algorithmic}
\usepackage{xcolor}

\usepackage{amsmath}

\usepackage{tikz}
\usepackage{enumerate}
\usepackage{booktabs}
\usepackage{makecell}
\usepackage{listings}
\usepackage{pifont}
\usepackage{algorithm}
\usepackage{array}
\usepackage[caption=false,font=footnotesize]{subfig}
\usepackage{textcomp}
\usepackage{stfloats}
\usepackage{url}
\usepackage{verbatim}
\usepackage{graphicx}

\usepackage{multirow}
\usepackage{colortbl}
\usepackage{tabularx}
\usepackage{graphicx}

\usepackage{multicol}
\usepackage{marvosym}
\newcommand{\Envelope}{\Letter}

\newcommand{\sys}{FlexServe}
\newcommand{\mem}{Flex-Mem}
\newcommand{\npu}{Flex-NPU}

\begin{document}

\date{}

\title{{\sys}: A Fast and Secure LLM Serving System for Mobile Devices with Flexible Resource Isolation}

\author{
\begin{tabular}{@{}c@{\hspace{0.9em}}c@{\hspace{0.9em}}c@{\hspace{0.9em}}c@{\hspace{0.9em}}c@{\hspace{0.9em}}c@{}}
{\rm Yinpeng Wu} & {\rm Yitong Chen} & {\rm Lixiang Wang} & {\rm Jinyu Gu} & {\rm Zhichao Hua}\textsuperscript{\Envelope} & {\rm Yubin Xia} 
\end{tabular}\\
{\rm Institute of Parallel and Distributed Systems, Shanghai Jiao Tong University}\\
{\rm \{wyp1536481268,yitongcheng,2042567212,gujinyu,xiayubin,zchua\}@sjtu.edu.cn}
}

\begin{abstract}

Device-side Large Language Models (LLMs) have grown explosively, offering stronger privacy and higher availability than their cloud-side counterparts.
During LLM inference, both the model weights and the user data are valuable, and attackers may compromise the OS kernel to steal them.
ARM TrustZone is the de facto hardware-based isolation technology on mobile devices, used to protect sensitive applications from a compromised OS.
However, protecting LLM inference with TrustZone incurs significant overhead to both the secure inference and the normal aplications, due to two challenges: the inflexible resource isolation and the inefficient secure resource management.

To address these challenges, this paper presents {\sys}, a fast and secure LLM inference system for mobile devices.
The key idea is to decouple the access permission from the management permission of secure resources, so that the normal-world OS cannot access them but can still manage them as usual.
First, {\sys} introduces a Recallable Resource Isolation mechanism to construct Recallable Secure Memory ({\mem}) and a Recallable Secure NPU ({\npu}).
They can only be accessed by the secure world, but can be efficiently allocated and reclaimed by the normal-world OS.
Based on them, {\sys} further introduces a {\sys} Framework to run secure LLM inference in the secure world.
It works together with the normal-world OS to perform cooperative secure memory management.
We implement a prototype of {\sys} and compare it with two TrustZone-based strawman designs.
The results show that {\sys} achieves average TTFT speedups of 10.05$\times$ over the strawman and 2.44$\times$ over an optimized strawman.

\end{abstract}

\maketitle


\section{Introduction}

Device-side Large Language Models (LLMs) have grown explosively in recent years~\cite{dubey2024llama,bai2023qwen,abdin2024phi, team2025gemma, xu2024device,li2024large}, offering stronger privacy and higher availability than their cloud-side counterparts.
Moreover, device-side LLMs can be trained or fine-tuned on domain-specific datasets, making them well suited to specialized tasks~\cite{fingpt, li2024ferret, yang2025ui, you2024ferret, xiao2021lawformer, xie2023pixiu}, and mobile AI applications can achieve higher intelligence by orchestrating multiple LLMs~\cite{zhang2025mobiagent, autogpt, babyapi, li2025apple, li2023robust,zhang2024you}.
Model vendors have released small-scale models tailored for mobile devices~\cite{dubey2024llama,bai2023qwen,abdin2024phi, team2025gemma}, and developers are increasingly integrating device-side LLMs into their applications~\cite{zhang2025mobiagent,apple-ai,li2025apple,google-llm-1,replika,notion,samsung-ai}.

Deploying LLMs on mobile devices introduces new security challenges.
The LLM model itself is highly valuable, often costing millions of dollars to train~\cite{kandpal2025position, cottier2024rising}.
The LLM service may process a wide range of sensitive data on mobile devices, including chat history, screen content, and more.
These factors make LLM inference an attractive target for attackers.
Since the existing OS kernel is large and prone to bugs~\cite{linux-code, linux-cve}, attackers may compromise the kernel to steal model weights or user data during LLM inference.

\begin{figure}[tb]
    \centering
    \subfloat[]{
        \includegraphics[width=0.46\linewidth]{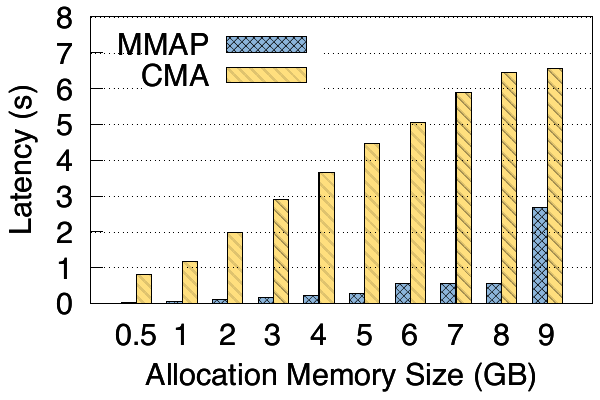}
        \label{fig:cma-diff-alloc}
    }
    \subfloat[]{
        \includegraphics[width=0.46\linewidth]{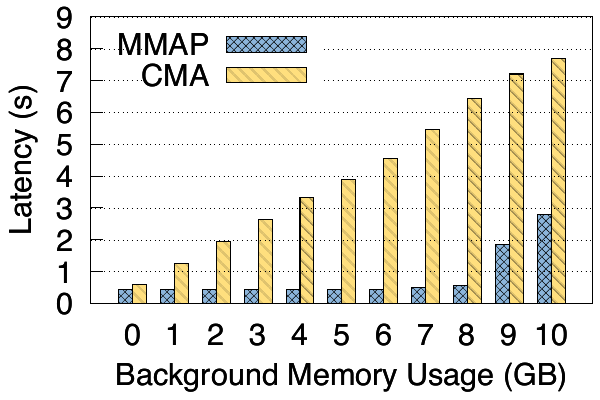}
        \label{fig:cma-diff-back}
    }
    \vspace{-0.3cm}
    \setlength{\belowcaptionskip}{-5pt} 
    \caption{Latency of allocating different sizes of memory (a) and of allocating 8GB memory under different background memory loads (b).
    }
    \label{fig:motiv-cma-lat}
\end{figure}

\begin{figure}[tb]
    \centering
    \includegraphics[width=\linewidth]{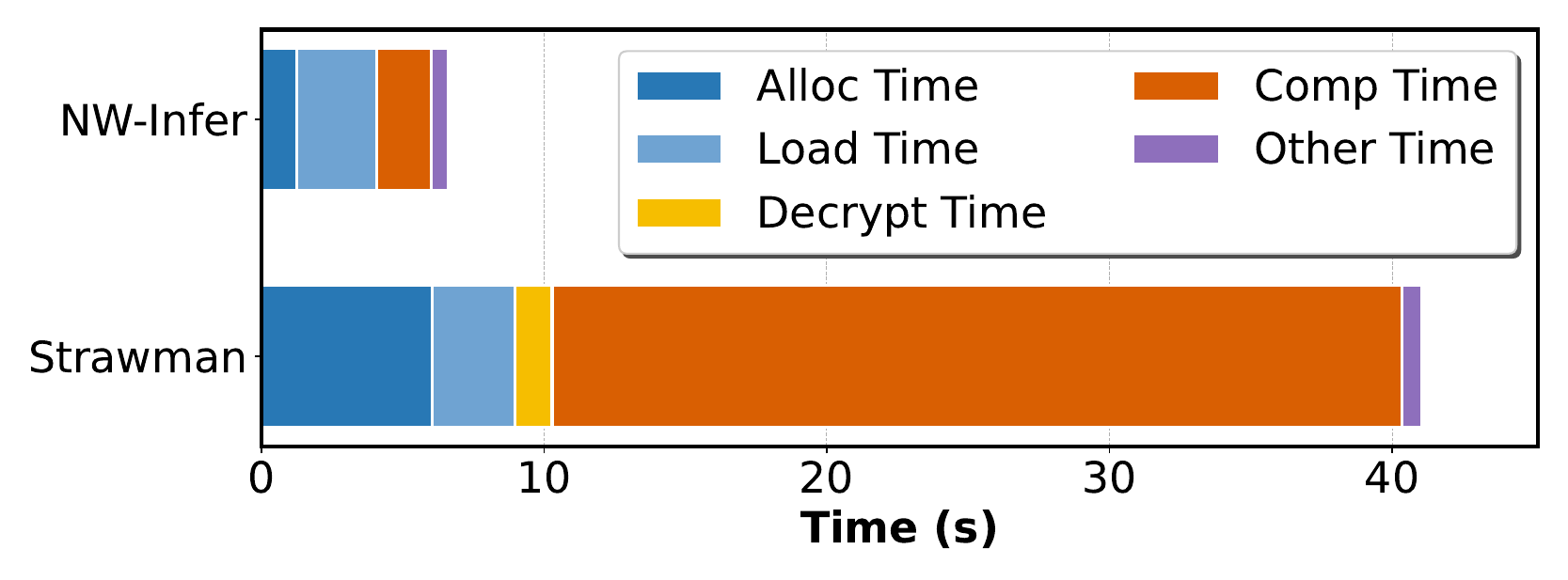}
    \vspace{-0.5cm}

    \setlength{\belowcaptionskip}{-5pt} 
    \caption{Breakdown of the TTFTs of the normal-world inference and the TrustZone-based strawman (Llama3.1 8B with a 128-token prompt).}
    \label{fig:motiv-breakdown}
\end{figure}

ARM TrustZone~\cite{alves2004trustzone} is a hardware isolation technology that protects sensitive applications from a compromised OS on mobile devices~\cite{santos2014using, guan2017trustshadow, li2014building, li2015adattester,luo2018tz}.
It provides a Trusted Execution Environment (TEE) called the secure world.
A strawman approach is to run LLM inference directly in the secure world.
However, this approach degrades the performance of both the LLM inference and the normal-world applications, because of the following two challenges:

\textbf{Challenge-1: Inflexible Secure Resource Isolation.}
TrustZone's secure resource isolation mechanism, including both secure memory and secure devices, is inflexible.
Neither the memory nor the NPU can be efficiently switched between the normal and secure worlds.
For memory, TrustZone protects secure memory using a limited number of regions, each of which must be physically contiguous.
LLM inference, however, requires a large amount of secure memory (GB level).
Allocating such a large contiguous memory region is slow, especially on memory-constrained mobile devices where most memory is already occupied by cached data and code of commonly-used applications~\cite{android-memory, android-memory-2, android-memory-3}.
As shown in Figure~\ref{fig:motiv-cma-lat}, allocating 8GB of contiguous memory takes about 6.44s, far longer than the prefill time of LLM inference, whereas a normal \emph{mmap} takes only 0.56s.
For devices, although the NPU can be dynamically switched between the normal and secure worlds, the internal state of the NPU driver cannot be switched.
Consequently, existing mobile devices always configure the NPU as a normal-world device, leaving it unusable in the secure world.
Relying solely on the CPU for LLM inference significantly degrades performance, as shown in Figure~\ref{fig:motiv-breakdown}.

\textbf{Challenge-2: Inefficient Secure Resource Management.}
On mobile devices, the OS caches the code and data of different applications~\cite{android-memory, android-memory-2, android-memory-3} and schedules memory among them to deliver the best user experience~\cite{linux_kswapd_physical_memory,android_lmkd}.
The secure LLM inference framework also needs to cache model weights and KV caches to improve performance.
However, once memory is configured as secure, neither normal nor secure world has a complete view of the other's workload.
As a result, the system cannot achieve optimal performance for both the normal-world applications and the secure LLM inference.
Furthermore, state-of-the-art (SOTA) mobile agent applications~\cite{zhang2025mobiagent, autogpt, babyapi, li2025apple, li2023robust,zhang2024you} employ multiple LLMs to achieve higher intelligence, further complicating secure memory management.


To address the above two challenges, this paper presents {\sys}, a fast and secure LLM inference system for mobile devices.
The key idea is to decouple the access permission from the management permission, which prevents the normal-world OS from accessing the secure resources while still allowing it to manage them as usual.

\emph{For the challenge of inflexible resource isolation}, {\sys} proposes a Recallable Resource Isolation mechanism to construct Recallable Secure Memory ({\mem}) and a Recallable Secure NPU ({\npu}).
Leveraging the Stage-2 Page Table (S2PT), {\sys} ensures that {\mem} and {\npu} can only be accessed by the secure world, yet remain efficiently allocatable and reclaimable by the normal-world OS.
Switching memory and the NPU between unprotected and protected modes is fast.
An on-demand protection mechanism further eliminates the virtualization overhead when no secure inference tasks are active.

\emph{For the challenge of inefficient secure resource management}, {\sys} introduces a {\sys} Framework that runs the secure LLM inference in TrustZone's secure world.
It reuses the existing secure-world software stack to execute inference, and leverages {\mem} and {\npu} to protect runtime data and to accelerate computation.
It cooperates with the normal-world OS to perform cooperative secure memory management.
The normal-world OS monitors the memory pressure to schedule memory between normal-world applications and the secure LLM inference, while the {\sys} Framework performs LLM-aware caching and memory reclamation.
This cooperative memory management achieves both high secure inference performance and low normal-world application overhead.

We implemented a prototype of {\sys} on a NanoPC-T6 development board~\cite{friendlyelec-nanopc-t6} with 8-core CPU and a 6TOPS NPU.
We compare {\sys} against two TrustZone-based strawman designs.
The results show that {\sys} achieves an average \textbf{10.05$\times$} TTFT speedup over the strawman, and an average \textbf{2.44$\times$} TTFT speedup over an optimized strawman with pipelining and the secure NPU enabled.
For agent workflows, the end-to-end speedup reaches up to \textbf{24.30$\times$} and \textbf{4.05$\times$} over the strawman and the optimized strawman, respectively.
While achieving high inference performance, {\sys} keeps normal-world applications with relative performance of \textbf{97.2}\% compared with the non-inference baseline.
Our main contributions are:
\begin{itemize}
    \item The Recallable Resource Isolation mechanism, which constructs {\mem} and {\npu} that can only be accessed by the secure world but can be allocated and reclaimed by the normal-world OS efficiently.
    
    \item The {\sys} Framework, which runs in the secure world and leverages {\mem} and {\npu} to perform secure LLM inference.
    It provides LLM-aware memory reclamation and dynamic caching, cooperating with the normal-world OS to perform efficient memory management.
    
    \item A prototype implementation of {\sys} and a detailed evaluation with it.
    The results show that {\sys} achieves an average \textbf{$10.05\times$} TTFT speedup compared to the TrustZone-based strawman, while preserving the performance of the co-running normal-world applications (\textbf{97.2}\% of the non-inference baseline).

\end{itemize}

\section{Background and Motivation}
\label{sec:motiv}

\subsection{LLMs on Mobile Device}
\label{subsec:motiv-llm}

Device-side LLMs~\cite{dubey2024llama,bai2023qwen,abdin2024phi, team2025gemma, xu2024device,li2024large} are deployed directly on mobile devices, so that users no longer need to upload their data to the cloud, which reduces the risk of data leakage and removes the dependency on network connectivity.
Model vendors have released small-scale models suitable for mobile deployment, such as Llama~\cite{dubey2024llama}, Phi-4~\cite{abdin2024phi}, Qwen3~\cite{bai2023qwen} and Gemma~\cite{team2025gemma}, and developers are increasingly integrating device-side LLMs into their applications~\cite{zhang2025mobiagent,apple-ai,li2025apple,google-llm-1,replika,notion,samsung-ai}.

\textbf{Multiple Models in One Device}:
Unlike cloud-side LLMs, which are large in size and provide general intelligence, device-side LLMs are small-scale models that can be trained or fine-tuned on specific datasets, making them well suited to specific tasks such as financial analysis, UI navigation, and more~\cite{fingpt, li2024ferret, yang2025ui, you2024ferret, xiao2021lawformer, xie2023pixiu}.
State-of-the-art mobile agent applications also employ multiple LLMs to achieve high intelligence~\cite{zhang2025mobiagent, autogpt, babyapi, li2025apple, li2023robust,zhang2024you}.

\textbf{Security Challenges of Device-side LLMs}:
First, the LLM model weights are valuable assets, as they are trained with a large amount of data and computational resources~\cite{kandpal2025position, cottier2024rising}.
Second, AI applications feed various types of data, including chat history and screen content, as input to the LLM.
Compromising the LLM inference system can therefore leak a significant amount of sensitive user information.
Unfortunately, existing mobile OS kernels have a large code base and are prone to bugs, which makes them vulnerable to attacks.
For example, Android relies on the Linux kernel, which comprises 40 million lines of code~\cite{linux-code} and has 9,756 reported CVEs~\cite{linux-cve}.
Attackers can compromise the kernel to steal the model weights and user data during the LLM inference procedure.

\subsection{Challenges of Protecting LLM with TrustZone}
\label{subsec:motiv-challenges}

\textbf{ARM TrustZone}:
ARM TrustZone~\cite{alves2004trustzone} is a hardware security extension that divides the processor into a normal world and a secure world.
All hardware resources, including memory and devices, can be partitioned into normal and secure modes.
The normal world cannot access secure resources, whereas the secure world can access all resources.
The commodity OS and normal applications run in the normal world.
The secure world is a Trusted Execution Environment (TEE), which hosts secure applications.
Even a malicious OS kernel cannot compromise the confidentiality or integrity of secure-world applications.
TrustZone has been widely used to protect various applications~\cite{santos2014using, guan2017trustshadow, li2014building, li2015adattester,luo2018tz}.

To protect LLM inference from an untrusted OS, a strawman approach is to run it in the secure world of TrustZone.
However, this approach faces two main challenges that significantly degrade inference performance.

\textbf{Challenge-1: Inflexible Secure Resource Isolation.}
Switching resources between the normal and secure worlds is inflexible.
For physical memory, TrustZone can configure only a limited number (e.g., 8) of physically contiguous regions as secure.
Allocating a secure memory region therefore requires merging fragmented free pages into a contiguous region, which is slow.
For devices, although they can be dynamically configured as normal or secure, the driver manages the status of each device, and switching the driver status is complex.
As a result, existing mobile devices choose to statically partition the hardware resources, configuring only a limited amount of physical memory and security-related devices as secure during system boot.
This suffices for traditional secure applications, such as key management and kernel integrity protection~\cite{santos2014using, guan2017trustshadow, li2014building, li2015adattester,luo2018tz}.

Unfortunately, LLM inference demands a large amount of memory and the NPU.
A Llama3.1 8B model with 8-bit quantization requires 7.5GB of memory for the model weights.
Statically partitioning 8GB secure memory significantly hurts the performance of normal-world applications.
Dynamically allocating 8GB secure memory, which must be physically contiguous, is very slow.
We evaluate the allocation latency of the Linux Contiguous Memory Allocator (CMA) on NanoPC-T6~\cite{friendlyelec-nanopc-t6} with 16GB memory.
The CMA takes \textbf{6.44s} to allocate 8GB of memory, whereas a normal \emph{mmap} takes only \textbf{0.56s}, under 8GB of background memory usage, which is not high for a modern 16GB mobile device (Figure~\ref{fig:motiv-cma-lat}).
On the other hand, if the NPU is inaccessible from the secure world, LLM inference becomes much slower.
Figure~\ref{fig:motiv-breakdown} breaks down the time to first token (TTFT) of a Llama3.1 8B model with a prompt length of 128.
The computation time of the strawman (using the CPU) is \textbf{30.06s}, whereas the normal-world inference (using the NPU) takes only \textbf{1.94s}.

\textbf{Challenge-2: Inefficient Secure Resource Management.}
Mobile OSes such as Android follow the principle that ``free memory is wasted memory'' and use available memory for caching to improve the user experience~\cite{android-memory, android-memory-2, android-memory-3}.
In our evaluation, on a OnePlus 12 running ColorOS 16.0.3 (based on Android 16) with 16GB of memory, \textbf{8.83GB} of memory is in use immediately after system boot.
The OS continuously monitors memory pressure and schedules memory among applications to achieve the best user experience.
Secure LLM inference likewise needs to cache model weights and KV caches to improve performance.

However, once memory is configured as secure, it is managed by the secure-world OS rather than the normal-world OS.
Neither side has a complete view of the other's workload.
As a result, the system cannot efficiently schedule memory between normal-world applications and the secure LLM inference.
In our evaluation, if the secure-world OS retains 8GB secure memory as an unreclaimable cache to accelerate LLM inference, the PostgreSQL~\cite{postgresql} throughput (measured with \texttt{sysbench}~\cite{sysbench}) drops from \textbf{1545.51} TPS to \textbf{882.81} TPS.
Conversely, if the secure-world OS releases the secure memory after each inference, the TTFT increases by \textbf{$2.54\times$} for Llama3.1 8B with a sequence length of 128.
More results are presented in Section~\ref{subsec:eval-normal}.
This separation of memory management makes it difficult to achieve optimal performance for both the normal-world applications and the secure LLM inference.

\subsection{ARM Virtualization}
\label{subsec:motiv-hyper}

The ARM virtualization extension supports running Virtual Machines (VMs) on the ARM platform.
It introduces a hypervisor mode (EL2) for the hypervisor, which manages resources and traps critical operations from VMs.
A two-stage address translation mechanism is introduced to support memory virtualization.
The Stage-1 Page Table (S1PT), controlled by the OS kernel, translates the virtual address (VA) to the intermediate physical address (IPA) for each VM.
The Stage-2 Page Table (S2PT), controlled by the hypervisor, then translates the IPA to the physical address (PA) for each VM.
The System MMU (SMMU) is introduced to enforce access control for DMA operations.
{\sys} leverages the virtualization extension to implement Recallable Resource Isolation.

\section{Overview}
\label{sec:overview}

\begin{figure*}[tb]
    \centering
    \includegraphics[width=\linewidth]{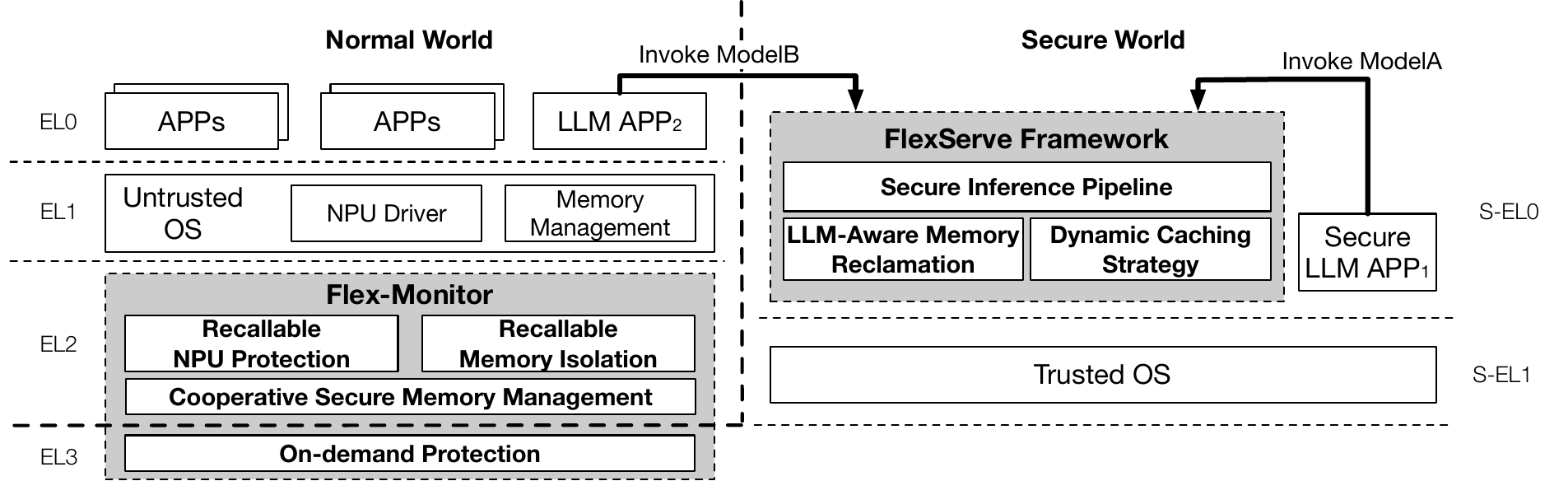}
    \setlength{\abovecaptionskip}{-2pt}
    \caption{\textbf{System overview of {\sys}}: The Flex-Monitor constructs the {\mem} and {\npu}, and the {\sys} Framework provides a fast and secure LLM inference framework.}
    \label{fig:system-overview}
\end{figure*}

\subsection{Design Goals}
\label{subsec:overview-goals}

{\sys} aims to provide a fast and secure LLM inference system for mobile devices.
The detailed goals are:

\begin{itemize}

    \item \textbf{Security}: 
    The confidentiality and integrity of model weights and input/output are protected during LLM inference against a compromised OS kernel.
    
    \item \textbf{High Inference Performance}:
    Both the Time to First Token (TTFT) and the Time Between Tokens (TBT) should be minimized.
    High performance is maintained when different models are invoked, especially for multi-model agent workflows.
    
    \item \textbf{Low Impact to Unprotected Applications}:
    The performance overhead to normal-world applications should be minimized.

    

\end{itemize}

\subsection{Threat Model}
\label{subsec:overview-threat}

{\sys} aims to protect LLM inference from attackers with kernel privileges.
Both the confidentiality and the integrity of the model weights and the input/output are protected.
All normal-world applications are considered untrusted.
{\sys} assumes that the initial kernel code is benign and that secure boot protects its integrity.
However, the kernel may contain bugs and could be compromised after system boot.
Consequently, with kernel privileges an attacker could access or modify arbitrary memory pages or peripherals to compromise the LLM inference system.
The secure-world components and the lightweight Flex-Monitor are trusted.

{\sys} handles requests from both normal-world and secure-world clients.
As with existing TrustZone-based systems, {\sys} cannot prevent attackers from directly stealing or tampering with the input/output of normal-world clients.
Nevertheless, the model weights remain protected.
Side-channel attacks, physical attacks, and DoS attacks are out of scope.
Section~\ref{subsec:discuss-security} presents a detailed security analysis.

\subsection{System Overview}
\label{subsec:overview-system}

This paper presents {\sys}, a fast and secure LLM inference system for mobile devices.
{\sys} runs LLM inference in the secure world of TrustZone, and solves the challenges of inflexible resource isolation and inefficient resource management (Section~\ref{subsec:motiv-challenges}).
The main idea is to decouple the access permission from the management permission, which prevents the normal-world OS from accessing the secure resources while still allowing it to manage them as usual.
To achieve this, {\sys} first introduces a new Recallable Resource Isolation mechanism to construct Recallable Secure Memory ({\mem}) and Recallable Secure NPU ({\npu}).
They cannot be accessed by the normal-world OS, yet can be efficiently allocated and reclaimed by it, which solves the first challenge of inflexible secure resource isolation.
Building on {\mem} and {\npu}, {\sys} further introduces the {\sys} Framework to run the secure LLM inference in TrustZone's secure world. 
It cooperates with the normal-world OS to perform the LLM-aware secure memory management and accelerate the inference, which solves the second challenge of inefficient secure resource management.
Based on them, {\sys} can protect both the privacy and integrity of the LLM inference, while achieving high inference performance and low overhead to normal-world applications.

Figure~\ref{fig:system-overview} shows the detailed design of {\sys}.
It uses a Flex-Monitor, running across normal EL2 (hypervisor mode) and secure EL3 (monitor mode), to provide the Recallable Resource Isolation.
Flex-Monitor first constructs the {\mem}, a page-granular secure memory abstraction, which can be allocated and recalled efficiently (Section~\ref{subsec:rs-mem}).
It introduces a cooperative secure memory management to allow existing normal-world memory management service to manage both the normal memory and the {\mem} (Section~\ref{subsec:rs-manage}).
The Flex-Monitor then constructs the {\npu}, a secure NPU abstraction (Section~\ref{subsec:rs-npu}).
Instead of using two NPU drivers and switching status between them, {\sys} reuses the normal-world NPU driver to control the {\npu}, thereby minimizing the Trusted Computing Base (TCB).
The Flex-Monitor leverages two-stage address translation to protect both the {\mem} and the {\npu}.
An on-demand protection mechanism is introduced to eliminate this virtualization overhead when no secure inference tasks are active (Section~\ref{subsec:rs-ondemand}).


Based on {\mem} and {\npu}, {\sys} further constructs the {\sys} Framework within TrustZone's secure world.
It reuses the existing secure-world software stack to execute the inference framework, and leverages {\mem} and {\npu} to protect runtime data and accelerate inference.
A secure inference pipeline is introduced to hide the latency overhead of secure loading and cryptographic operations (Section~\ref{subsec:framework-pipeline}).
Benefiting from the page-granular and flexible protection of {\mem}, a dynamic caching strategy is introduced to cache the model weights and KV cache in the {\mem} (Section~\ref{subsec:framework-cache}).
Then, the LLM-aware memory reclamation is introduced to decide which pages to reclaim when normal-world OS requests memory reclamation (Section~\ref{subsec:framework-reclaim}).
It also re-schedules the cache to the optimal distribution after the reclamation.
The {\sys} Framework can handle requests from both normal-world and secure-world applications.
The lifecycle of {\sys} is detailed in Section~\ref{subsec:framework-lifecycle}.

\section{Recallable Resource Isolation}
\label{sec:rs}

\subsection{Recallable Secure Memory}
\label{subsec:rs-mem}

\begin{figure}[tb]
    \centering
    \includegraphics[width=\linewidth]{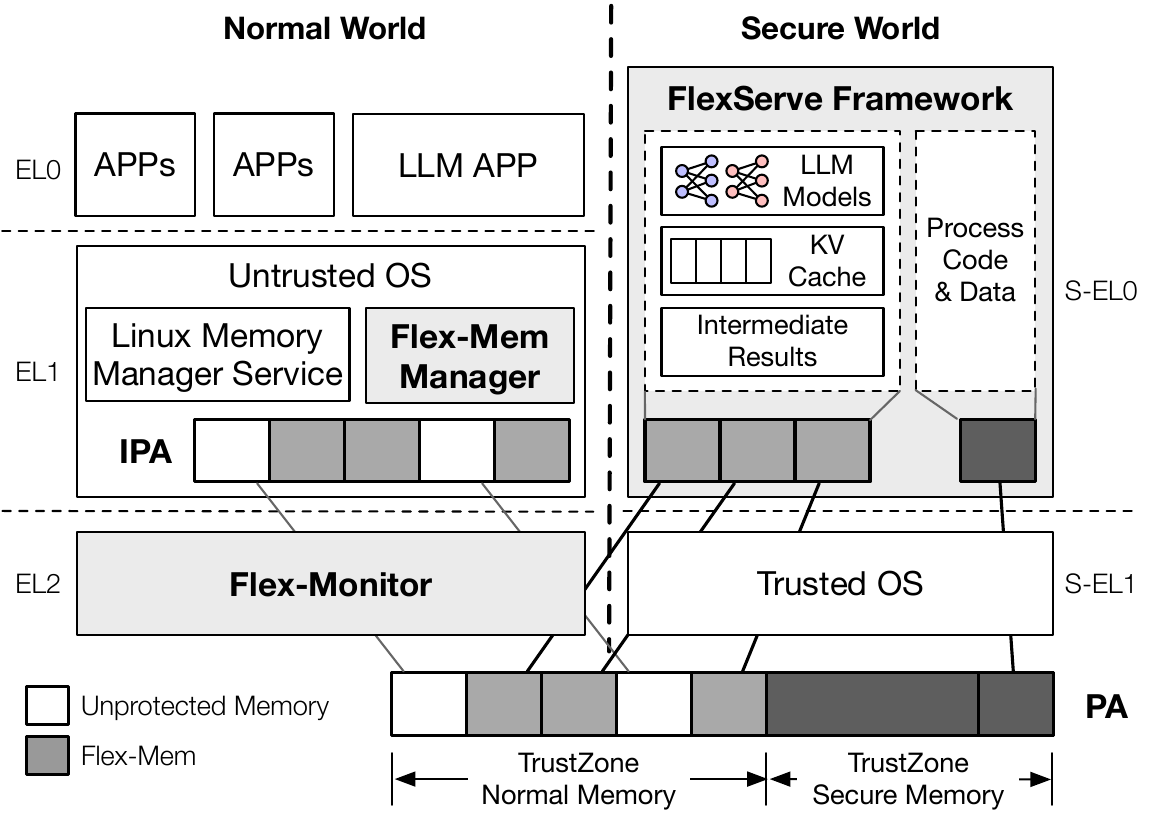}
    \setlength{\abovecaptionskip}{-3pt}
    \caption{Memory Protection of {\sys}.}
    \setlength{\belowcaptionskip}{0pt}
    \label{fig:rs-memory}
\end{figure}

{\sys} divides all memory resources into three types: unprotected memory, {\mem}, and TrustZone's secure memory (Figure~\ref{fig:rs-memory}).
Both unprotected memory and {\mem} are the normal memory of TrustZone.
Unprotected memory is used by the untrusted OS and applications.
The {\mem} is a recallable, page-granular secure-memory abstraction designed for secure LLM inference.
Any unprotected memory page can be switched to a {\mem} page, which can later be returned to unprotected memory when memory pressure rises or the {\mem} is unused.

As shown in Figure~\ref{fig:rs-memory}, the Flex-Monitor, running at EL2, isolates all {\mem} from the untrusted OS kernel by leveraging the Stage-2 Page Table (S2PT).
For each {\mem} page, the Flex-Monitor removes the IPA-to-PA mapping of that page from the normal-world S2PT.
Therefore, neither the untrusted OS nor its applications can access it.
After allocation, the secure-world Trusted OS maps the {\mem} pages into the {\sys} Framework's address space.
Details about the {\sys} Framework are provided in Section~\ref{sec:framework}.
If a {\mem} page is reclaimed, the Flex-Monitor remaps it in the normal-world S2PT and returns ownership to the normal-world OS.


\textbf{DMA Protection:}
Direct Memory Access (DMA) may be abused to access {\mem} pages.
The Flex-Monitor prevents this by removing DMA mappings for {\mem} pages from the SMMU page tables, which perform address translation for DMA operations.
The Flex-Monitor first unmaps the SMMU's MMIO region in the S2PT, so that any updates to SMMU configuration registers trap to the Flex-Monitor.
It then tracks or validates active SMMU page-table roots and removes mappings for {\mem} pages.
Since the SMMU base address registers are rarely accessed, the hooking overhead is minimal.
The on-demand protection mechanism can further reduce this protection overhead (Section~\ref{subsec:rs-ondemand}).

\subsection{Cooperative Secure Memory Management}
\label{subsec:rs-manage}

The Flex-Monitor provides allocation and reclamation interfaces for {\mem}, and leverages a kernel-level {\mem} manager to manage it.

\textbf{{\mem} Allocation}:
The Flex-Monitor provides the \emph{rsmem\_alloc(size)} interface to allocate {\mem} pages.
The {\sys} Framework invokes this interface to request additional {\mem} pages.
The Flex-Monitor then asks the {\mem} manager to allocate physical pages from the normal-world OS.
These pages are unmapped from the normal-world S2PT and marked as {\mem} pages.
Finally, the Flex-Monitor returns the allocated {\mem} pages to the Trusted OS, which maps them into the {\sys} Framework's address space.

\textbf{{\mem} Reclamation}:
The Flex-Monitor provides the \emph{rsmem\_reclaim(size)} interface to reclaim {\mem} pages.
The normal-world OS invokes this interface to request reclamation of {\mem}.
The Flex-Monitor then asks the {\sys} Framework to select {\mem} pages for reclamation.
The framework provides LLM-aware memory reclamation; details are given in Section~\ref{subsec:framework-reclaim}.
{\mem} pages containing model weights only need to be zeroed.
Pages containing KV caches must be encrypted and written back to storage before being zeroed.
After these steps, the Flex-Monitor remaps the pages in the normal-world S2PT.
Finally, these pages are marked as normal memory and returned to the normal-world OS.



\textbf{Asynchronous Reclamation}:
We observe that there is a time window between memory reclamation and the actual use of the reclaimed pages.
For example, when the OS detects memory pressure, it reclaims memory pages and remaps them to an application.
However, the application may not access all of these pages immediately.
Based on this observation, {\sys} introduces an asynchronous reclamation mechanism that returns to the normal-world OS immediately after selecting the pages to reclaim.
These pages are marked as claimed so that the normal-world OS can reuse them.
All remaining operations, including encryption, writing back to storage, zeroing, and remapping, are performed asynchronously.
If a subsequent access to a page occurs before the asynchronous operations complete, an S2PT page fault is triggered.
The access is then blocked until the operations finish.
The asynchronous reclamation mechanism significantly reduces reclamation latency, especially when a large number of pages must be reclaimed (details in Table~\ref{table:micro-bench}).


\textbf{Integration with Normal-World Memory Management}:
{\sys} allows {\mem} to be managed by the normal world's existing memory management services.
The Linux kernel provides memory watermark mechanism: when free memory falls below a low watermark, \texttt{kswapd}~\cite{linux_kswapd_physical_memory} performs asynchronous memory reclamation, and subsequent allocation failures may trigger direct memory reclamation.
Android introduces \emph{lmkd}~\cite{android_lmkd} to detect memory pressure via Linux Pressure Stall Information (PSI)~\cite{linux_psi} and to perform memory reclamation accordingly.
The \emph{rsmem\_reclaim} interface is integrated into these memory management services.
Meanwhile, {\sys} also uses the {\mem} manager to monitor memory pressure.
When the manager detects low memory pressure, it invokes \emph{rsmem\_alloc} to allocate {\mem} pages for the {\sys} Framework.
Through this cooperative memory management, the normal-world OS decides whether to reclaim or allocate {\mem} pages, while the {\sys} Framework decides which secure pages to return or reload.



\subsection{Recallable NPU Protection}
\label{subsec:rs-npu}

{\sys} introduces the Recallable Secure NPU ({\npu}) to efficiently enable the NPU for secure LLM inference.
{\npu} operates on a time-multiplexing model: the NPU is either in unprotected mode, accessible by the normal world, or switched into {\npu} mode, where it is exclusively available to the secure world.

When the NPU is in {\npu} mode, the Flex-Monitor prevents the normal-world OS from accessing it.
The ARM architecture uses Memory-Mapped I/O (MMIO) to access devices, including the NPU.
Therefore, the Flex-Monitor removes the NPU's MMIO region from the normal-world's S2PT, effectively blocking kernel access.

Subsequently, {\sys} reuses the normal-world NPU driver to control the {\npu}.
The Flex-Monitor constructs an isolated {\npu} sandbox to protect the NPU driver when the NPU is in {\npu} mode.
Specifically, the Flex-Monitor maintains an additional S2PT for this sandbox.
When the secure world invokes the protected NPU driver, the Flex-Monitor switches to the sandbox's S2PT.
The NPU's MMIO region is mapped within the sandbox's S2PT, allowing the driver to access the NPU.
Both the driver's code and data are mapped in the sandbox's S2PT but unmapped from the normal-world OS's S2PT.
This prevents the OS from tampering with the protected NPU driver's code and data.
Although the driver retains residual state from unprotected mode, NPU task launching is a stateless operation.
Thus, the remaining state does not influence {\npu} task execution.

Furthermore, the Flex-Monitor enforces that the {\npu} can access only {\mem} pages by restricting the {\npu}'s SMMU page-table mappings.
Unprotected memory pages are unmapped, ensuring that private data cannot be leaked to unprotected memory via the {\npu}.
Note that since different devices use distinct SMMU page tables, the Flex-Monitor still ensures that no other devices can access {\mem} pages.

When the {\npu} switches back to unprotected mode, the Flex-Monitor remaps the NPU's MMIO region and the driver's code and data into the kernel's S2PT.
The NPU's SMMU page table is also reverted to map only unprotected memory pages.

\subsection{On-demand Protection}
\label{subsec:rs-ondemand}

The Flex-Monitor leverages the S2PT to protect {\mem} and {\npu}, which may introduce performance overhead for normal-world applications.
On-demand protection is introduced to minimize this overhead.
This mechanism disables the protection when no secure inference task has been active for a specified time window, and re-enables it when a new task arrives.
The key challenge lies in preserving the integrity of the Flex-Monitor itself, as its code and data reside in normal memory.
Once the S2PT is disabled, the normal-world OS could potentially modify the Flex-Monitor and compromise the protection.

To address this, the Flex-Monitor is divided into an EL2 component and an EL3 component.
The EL2 component implements the main protection mechanisms, including Recallable Secure Memory and Recallable NPU Protection.
The EL3 component freezes the EL2 component to eliminate the virtualization overhead and to protect its integrity.
The EL3 component executes within TrustZone's secure memory.
It calculates and stores a hash of the EL2 component, covering both its code and data.
Subsequently, it disables the S2PT.
To re-enable the protection, the EL3 component restores the S2PT and verifies the integrity of the EL2 component against the stored hash.
The S2PT is also verified as part of the EL2 component's data.
The entire EL2 component is placed in a contiguous memory region to simplify hash calculation.
Note that the EL2 component does not contain any private data (e.g., model weights), so {\sys} only protects its integrity.

\section{{\sys} Framework}
\label{sec:framework}


The {\sys} Framework is implemented as a secure-world Trusted Application (TA).
It uses secure memory for its code and private runtime state, including global variables and the stack.
Inference data, including the model weights and the KV cache, is placed in {\mem}, while {\npu} accelerates computation.

\subsection{Secure Inference Pipeline}
\label{subsec:framework-pipeline}


Due to memory limitations, mobile devices cannot always keep the model weights resident in memory and must load them for each inference, causing the \emph{cold start} problem.
This issue is amplified in confidential inference, because the encrypted weights must also be decrypted.

{\sys} leverages pipeline parallelism to reduce the cold-start overhead by overlapping resource-disjoint steps in the prefill stage, which is partitioned into four steps:
1) \textbf{Memory allocation}: allocating memory for the model weights and the KV cache;
2) \textbf{Model loading}: loading the encrypted weights from storage;
3) \textbf{Model decryption}: decrypting the weights;
4) \textbf{Forward computation}: executing the prefill with {\npu} and CPU.
The decode stage can reuse the in-memory weights and KV cache.

The prefill stage processes a sequence of layers, where each layer depends only on the outputs of previous layers.
Within a layer, the steps are constrained only by their \emph{in-layer} dependency (allocate $\rightarrow$ load $\rightarrow$ decrypt $\rightarrow$ compute) and do not depend on the corresponding steps of earlier layers.
{\sys} therefore overlaps the allocation, loading, and decryption of layer $i{+}1$ with the computation of layer $i$.

\textbf{Pipeline Bottleneck}:
The overall latency is determined by the longest step.
Figure~\ref{fig:motiv-breakdown} breaks down the TTFT of an inference task with a prompt length of 128.
CMA-based memory allocation and CPU-based computation are the two bottlenecks.
{\sys} removes them using {\mem} and {\npu}.
Consequently, the prefill pipeline is primarily bottlenecked by model loading.
{\sys} further mitigates this bottleneck through the model-weight and KV cache caching strategy described in Section~\ref{subsec:framework-cache}.

\textbf{Enabling Inline Encryption/Decryption with {\sys}}:
In our evaluation, although the decryption time is small and not the pipeline bottleneck, it still introduces contention on the CPU and memory bandwidth, which slows down the computation step.
For a Llama3.1 8B model with a 128-token prompt, concurrent decryption increases the computation time from 2.825s to 3.332s, adding 17.85\% overhead.
{\sys} solves this problem using inline cryptographic hardware, such as the Qualcomm Inline Crypto Engine (ICE)~\cite{qualcomm_ice_fips140_2017} and Google UFS Inline Storage Encryption (ISE)~\cite{google_tensor_ufs_ise_fips140_2023}.
The Flex-Monitor dynamically isolates the inline cryptographic hardware for the {\sys} Framework by controlling the S2PT and IOMMU.
The isolation method is the same as that for {\npu} (Section~\ref{subsec:rs-npu}).
Using the inline cryptographic hardware, {\sys} decrypts the model weights and KV cache directly on the storage DMA path, eliminating the CPU and memory-bandwidth contention.


\subsection{Dynamic Caching Strategy}
\label{subsec:framework-cache}

The {\sys} Framework handles requests for different models, and both the model weights and the history KV caches can be cached.
The dynamic caching strategy traces all history requests to the {\sys} Framework and decides which model weights and KV caches to cache, based on the current {\mem} budget.

First, the {\sys} Framework decides which models to cache and assigns a cache budget to each of them.
The framework logs the invocation history of all models and uses the recent invocation rate (e.g., over the past hour) to decide which models to cache.
Only models whose invocation rate exceeds a user-defined threshold (e.g., 25\%) are selected.
If no model reaches this threshold, the top 3 models with the highest invocation rate are selected.
For each selected model, the per-model cache budget is determined by its 1) recent invocation rate, 2) recent invocation frequency, and 3) history prompt length.
A model with a higher invocation rate and frequency receives a larger cache budget.
In {\sys}'s secure pipeline, a long prompt length indicates that the computation step will be the bottleneck, and the I/O can be easily overlapped, so the model will receive a smaller cache budget.

Next, the framework decides how to cache the model weights and KV caches for each selected model.
The invocation rate is also used to select the history KV caches.
Due to the limited memory, each model's cache budget may be insufficient to cache even its model weights.
Therefore, {\sys} caches the earlier portion of the model weights together with the selected KV caches.

\textbf{Workflow-aware Cache Management}:
We observe that agent applications often follow several stable workflows~\cite{zhang2025mobiagent, autogpt, babyapi, li2025apple, li2023robust,zhang2024you}.
Based on this, the {\sys} Framework logs the history workflows, each of which contains a list of \emph{<model, prompt KV cache>} pairs.
For a new request, if the framework detects a matching workflow, the subsequent models and KV caches are prefetched in order.
The prefetch is pipelined with the decode stage of the previous model, so this latency can be hidden and subsequent models can warm-start.

\subsection{LLM-Aware Memory Reclamation}
\label{subsec:framework-reclaim}

The LLM-aware memory reclamation selects {\mem} pages and returns them to the Flex-Monitor for reclamation (Section~\ref{subsec:rs-manage}).
However, the reclamation priority differs from the caching priority.
Reclamation should select the pages with the lowest reclamation cost, whereas caching evicts the pages with the lowest reuse probability.
Specifically, reclaiming KV caches requires encrypting the pages and writing them back to storage, while reclaiming model weights only requires zeroing the pages.
{\sys} therefore provides a two-stage reclamation.

In the first stage, the framework reclaims the pages with the lowest reclamation cost.
Model-weight pages are reclaimed first, ordered by invocation rate from low to high, followed by KV cache pages.
The reclaimed pages are returned to the Flex-Monitor in the first stage.
The second stage is then performed asynchronously, reorganizing the cache according to the priority generated by the Dynamic Caching Strategy.
It may swap out some KV cache pages and reload the model-weight pages.

The {\sys} Framework also introduces asynchronous swapping to optimize the two-stage reclamation.
At runtime, when the workload is low, the framework encrypts the history KV caches and writes them back to storage.
These swapped-out KV cache pages also have a low reclamation cost and can be reclaimed in the first stage.
This reduces the cache reorganization latency and improves the cache efficiency.

\subsection{Lifecycle of {\sys}}
\label{subsec:framework-lifecycle}


\textbf{Secure Boot and System Initialization}:
With secure boot technology, the firmware can verify the integrity of the Flex-Monitor and the secure-world OS.
The {\sys} Framework is implemented as a TA, signed by the device vendor's private key.
When the {\sys} Framework is started, the secure-world OS loads and verifies the framework's binary, thereby ensuring its integrity.

\textbf{Secure Session and Handling Inference Requests}:
Before sending an inference request, the client application and the {\sys} Framework establish a secure channel.
It is implemented following the standard GlobalPlatform TEE specifications, by invoking the \texttt{TEEC\_OpenSession} function.
During this phase, the client can attest the framework to verify that 1) it runs in the secure world and 2) its integrity is guaranteed.
Note that, as with existing TrustZone-based systems, {\sys} cannot protect the normal-world client from untrusted normal-world OS.
However, {\sys} also allows another secure-world TA to invoke the {\sys} Framework.

After the secure channel is established, the client sends the inference request to the {\sys} Framework through the channel.
A secure inference request includes the input prompt and the model index.
The {\sys} Framework receives the request and runs the secure inference.
The output tokens are continuously generated and returned to the client through the secure channel.
After all outputs are returned, the {\sys} Framework generates a response proof and sends it to the client.

\section{Implementation}
\label{sec:impl}

We implement a prototype of {\sys} on a NanoPC-T6 development board~\cite{friendlyelec-nanopc-t6}.
The software stack is built upon Linux Kernel 6.1.57, and the secure world OS is OP-TEE 4.5.0\cite{optee_os}.

The system architecture comprises two primary components: Flex-Monitor and the {\sys} Framework.
Flex-Monitor consists of approximately 4.5K lines of code (LoC). 
It implements essential virtualization primitives, including S2PT management and SMMU configuration.
The {\sys} Framework is a lightweight LLM inference engine implemented in C/C++ as a Trusted Application (TA). 
Spanning 8.3K LoC, it currently supports Llama3 and Qwen3 series models.

The vendor's closed-source NPU runtime~\cite{rknn-llm} executes specified LLM models as a black box, preventing the integration of our framework's optimizations.
{\sys} overcomes this limitation by leveraging a community-driven reverse-engineering project~\cite{rk3588-npu} to perform INT8 matrix multiplication on the NPU. 
To approach the performance of the proprietary driver, {\sys} applies several engineering optimizations, including automatic matrix blocking, CPU/NPU pipelined block matrix GEMM.
Since the IOMMU only supports 4GB address space, {\sys} employs adaptive IOMMU page-table switching to allow the NPU to access the 8GB address space.

\section{Evaluation}
\label{sec:eval}

\subsection{Experimental Setup}
\label{subsec:eval-setup}

The evaluation aims to answer the following questions:
Q-1) What are the costs of the critical recallable-resource operations?
Q-2) How does {\sys} perform for inference without caching?
Q-3) How does {\sys} perform for multi-model and agent workflows with caching?
Q-4) How does {\sys}'s protection affect normal-world applications?
Q-5) How do {\sys}'s different optimizations contribute to the performance improvement?

{\sys} is compared with the following three baselines:
1) \textbf{NW-Base}: the unprotected inference system running in the normal world, with pipeline optimizations and NPU acceleration.
2) \textbf{Strawman}: the secure inference system running in TrustZone's secure world, which uses CMA to allocate secure memory and the CPU for secure computation.
3) \textbf{Strawman-OPT}: an optimized Strawman that additionally enables pipeline optimizations and the NPU.
It still uses CMA to allocate secure memory.

The evaluation is conducted on a NanoPC-T6 development board~\cite{friendlyelec-nanopc-t6}, which features an octa-core CPU, 16GB memory, a 6TOPS NPU, and an SSD with 2.65GB/s bandwidth.
The configuration is consistent with modern mobile devices.
Due to our test platform limitations, it does not support inline storage encryption/decryption. Therefore, all evaluations are conducted without this optimization.
By default, we use \emph{stress-ng}~\cite{stress-ng} to generate background memory load, e.g., 8GB.
It occupies memory without adding contention for CPU or memory bandwidth.
{\sys} targets mobile devices, where response latency is critical.
Accordingly, the evaluation primarily focuses on the Time to First Token (TTFT) for LLM inference and the response latency for agent applications.

\subsection{Micro-benchmarks}

\begin{table}[t]
    \centering
    \footnotesize
    
    \caption{Latency of critical operations (ms).}
    \label{table:micro-bench}
    
    \def\microBenchTopNumGap{0.75em}
    \def\microBenchBottomNumGap{0.75em}
    \def\microBenchSectionGap{0.15em}

    \begin{tabularx}{\linewidth}{@{}>{\raggedright\arraybackslash}X@{\hspace{\microBenchTopNumGap}}r@{\hspace{\microBenchTopNumGap}}r@{}}
    
        \toprule
		\textbf{{\mem} \& {\npu} Operations} & \textbf{FlexServe} & \textbf{Baseline} \\
        \midrule
        Memory Alloc (8GB)   & \textbf{568.58} & 6440.67 \\
        Memory Reclaim (8GB) & \textbf{157.69}  & 503.64 \\
        NPU SMMU Setup (8GB) & 435.48          & \textbf{429.74} \\
        NPU Task Launch      & 1.28            & \textbf{1.26}   \\
        {\npu} Mode Switch      & 0.21            & N/A   \\
        \midrule
    \end{tabularx}
    \vspace{\microBenchSectionGap}

    \begin{tabularx}{\linewidth}{@{}>{\raggedright\arraybackslash}X@{\hspace{\microBenchBottomNumGap}}r@{}}
		\multicolumn{2}{c}{\textbf{Critical {\sys} Operations}} \\
        \midrule
        Memory Reclaim w/o Async + Two-Stage (128MB) & 74.69\\ 
        Memory Reclaim w/o Async (128MB) & 12.21 \\ 
        Memory Reclaim (128MB) & 2.89 \\  
        S2PT Boot            & 0.13 \\
        Hash Check           & 2.83 \\
        File Load (8GB)      & 3265.34 \\
        Memory Decrypt (8GB, 4 Cores) & 1319.16 \\
        Prefill (Llama3.1 8B, 128 tokens) & 2825.11 \\
        \bottomrule
    \end{tabularx}
\end{table}

To answer question-1, we measure the latency of critical {\sys} operations.
\textbf{{\mem} \& {\npu} Operations}:
The results are shown in the top half of Table~\ref{table:micro-bench}.
For {\mem}, the baseline is CMA-based secure memory allocation and reclamation.
{\mem} is \textbf{11.33$\times$} and \textbf{3.19$\times$} faster than the CMA-based approach in memory allocation and reclamation, respectively.
For {\npu}, the baseline is unprotected NPU operations.
The mode-switch latency of {\npu} is minimal, at only \textbf{0.21ms}.
Compared with the unprotected baseline, {\sys} adds only \textbf{1.34\%} and \textbf{1.59\%} overhead to the NPU SMMU setup and task launching, respectively.

\textbf{Critical Operations of {\sys}}:
The results are shown in the bottom half of Table~\ref{table:micro-bench}.
For memory reclamation, asynchronous and two-stage optimizations can reduce the latency from \textbf{74.69ms} to \textbf{2.89ms} for 128MB memory.
S2PT boot and hash check are the two main components of the protection enabling step in the on-demand protection mechanism.
The total latency is \textbf{2.96ms}.

\subsection{Performance without Cache}

To answer question-2, we evaluate the prefill and decode performance of {\sys}.
The model weights and KV caches are disabled to reveal the cold-start performance.

\begin{figure}[t]
    \includegraphics[width=\linewidth]{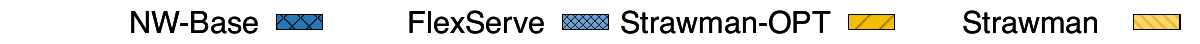}
    \vspace{-0.8cm}

    \centering
    \subfloat[Llama3.2 3B \label{fig:eval-ttft-model-1}]{
        \includegraphics[width=0.49\linewidth]{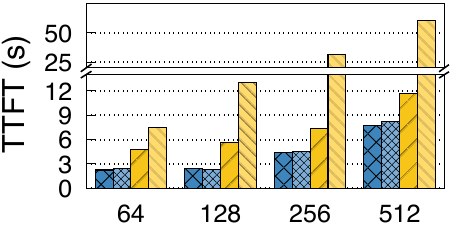}
    }
    \centering
    \subfloat[Llama3.1 8B \label{fig:eval-ttft-model-2}]{
        \includegraphics[width=0.49\linewidth]{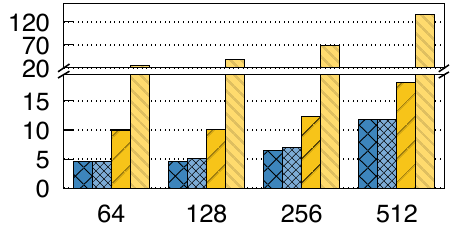}
    }
    \vspace{-0.3cm}

    \centering
    \subfloat[Qwen3 1.7B \label{fig:eval-ttft-model-3}]{
        \includegraphics[width=0.49\linewidth]{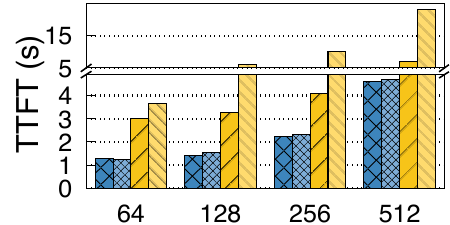}
    }
    \centering
    \subfloat[Qwen3 8B \label{fig:eval-ttft-model-4}]{
        \includegraphics[width=0.49\linewidth]{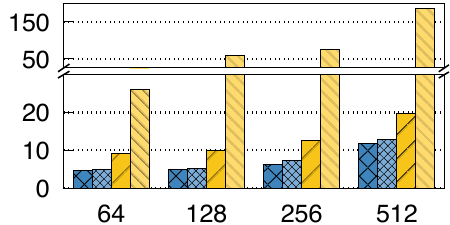}
    }

    \setlength{\belowcaptionskip}{-5pt} 
	\caption{Prefill time (TTFT) without cache.
    }
    \label{fig:eval-ttft-models}
\end{figure}

\textbf{Prefill Performance:}
We evaluate models ranging from 1.7B to 8B, all quantized to INT8 precision.
Figure~\ref{fig:eval-ttft-models} shows the TTFT for each model under different prompt lengths.
{\sys} achieves an average speedup of \textbf{7.15$\times$} and a maximum speedup of \textbf{14.44$\times$} over the Strawman.
Compared with Strawman-OPT, {\sys} delivers a speedup between \textbf{1.43$\times$} and \textbf{2.42$\times$} (\textbf{1.85$\times$} on average).
As the prompt length grows, the pipeline bottleneck shifts from I/O to computation, which narrows the speedup of {\sys}.
The evaluation platform does not feature a SOTA NPU; {\sys} would achieve a larger speedup with a faster NPU.

\begin{figure}[tb]
    \begin{minipage}[t]{\linewidth}
        \centering
        \includegraphics[width=\linewidth]{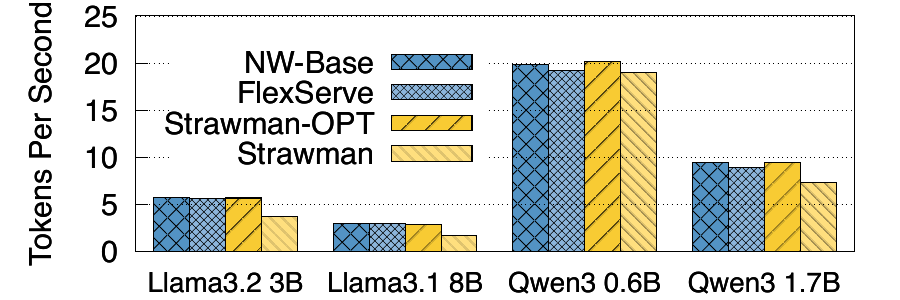}

        \setlength{\belowcaptionskip}{-5pt} 
        \caption{
            Decode throughput without cache.
        }        
        \label{fig:eval-decode}
    \end{minipage}
\end{figure}

\textbf{Decode Performance:}
{\sys} mainly targets the prefill phase, as it determines the response latency and is more important on mobile devices.
As shown in Figure~\ref{fig:eval-decode}, {\sys} improves the decode throughput by 24.14\% on average compared to the Strawman, owing to NPU acceleration.
{\sys}, Strawman-OPT, and NW-Base all achieve similar throughput because the model weights are already resident in memory and all three use pipelining and NPU.
Relative to the insecure NW-Base, {\sys} incurs a minor throughput reduction of 0.57\%--6.29\% (3.01\% on average).

\begin{figure*}[t]
    \centering
    \includegraphics[width=0.95\textwidth]{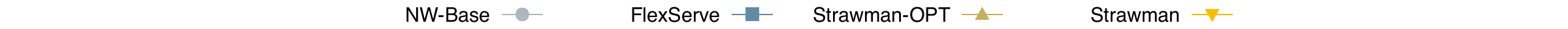}
    \vspace{-0.4cm}

    \centering
    \subfloat[Llama3.2 3B \label{fig:bg-sub1}]{
        \includegraphics[width=0.24\textwidth]{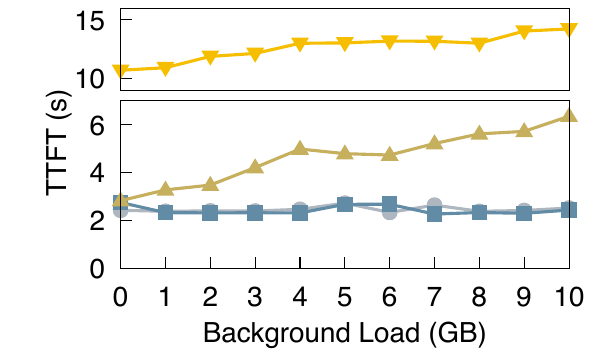}
    }
    \hfill
    \centering
    \subfloat[Llama3.1 8B \label{fig:bg-sub2}]{
        \includegraphics[width=0.24\textwidth]{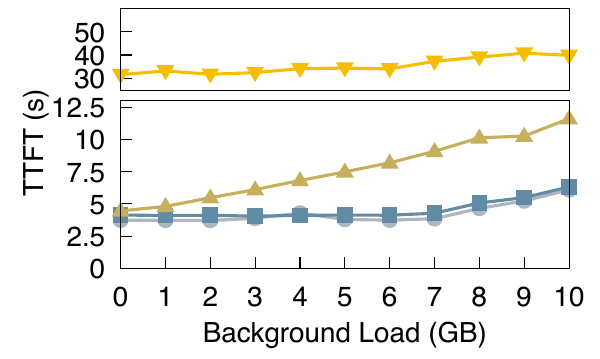}
    }
    \hfill
    \centering
    \subfloat[Qwen3 0.6B \label{fig:bg-sub3}]{
        \includegraphics[width=0.24\textwidth]{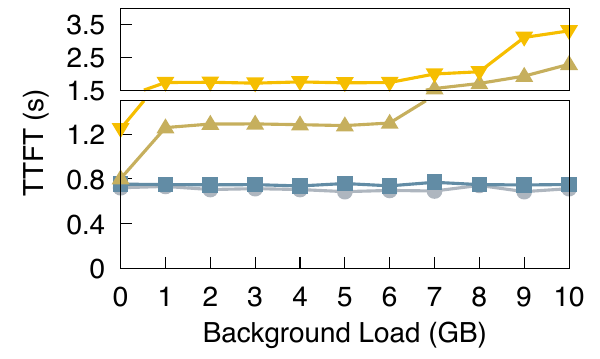}
    }
    \hfill
    \centering
	\subfloat[Qwen3 1.7B \label{fig:bg-sub4}]{
        \includegraphics[width=0.24\textwidth]{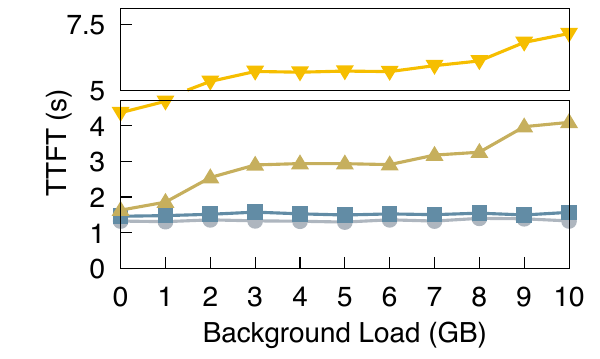}
    }

	\caption{
        Prefill time (TTFT) under varying background memory pressure, without cache.
    }
	\label{fig:diff-mem}
\end{figure*}

\begin{figure*}[t]
	\vspace{-0.3cm}
	
    \centering
    \includegraphics[width=1.0\textwidth]{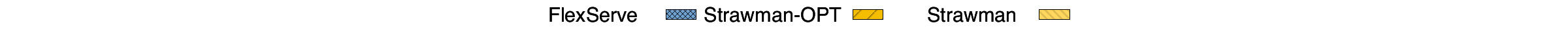}
    \vspace{-0.85cm}

    \centering
    \subfloat[Qwen3 1.7B + Llama3.1 8B \label{fig:mm-sub1}]{
        \includegraphics[width=0.24\textwidth]{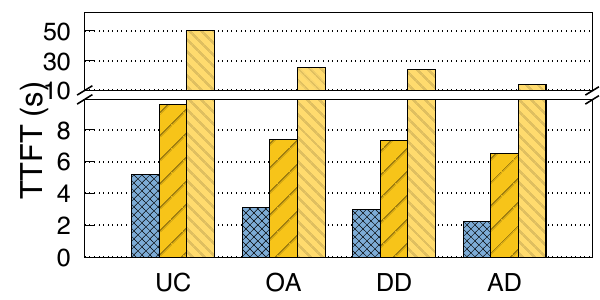}
    }
    \centering
    \subfloat[Qwen3 0.6B + Qwen3 1.7B \label{fig:mm-sub2}]{
        \includegraphics[width=0.24\textwidth]{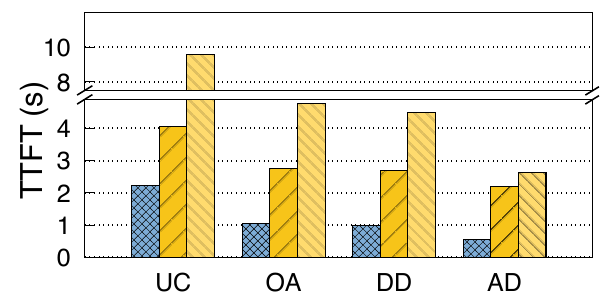}
    }
    \centering
    \subfloat[Qwen3 0.6B + Qwen3 8B \label{fig:mm-sub3}]{
        \includegraphics[width=0.24\textwidth]{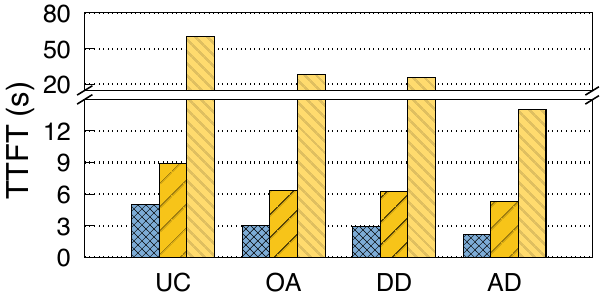}
    }
    \centering
	\subfloat[Qwen3 1.7B + Llama3.2 3B \label{fig:mm-sub4}]{
        \includegraphics[width=0.24\textwidth]{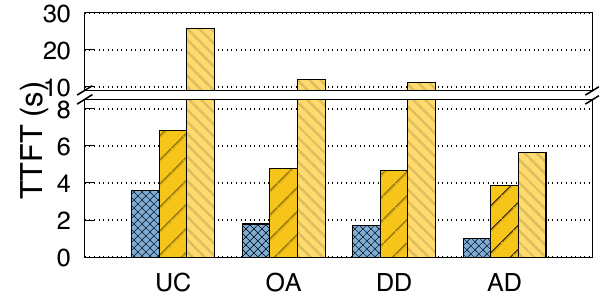}
    }
	\caption{
        Prefill time (TTFT) of different model groups and different benchmarks, with 4GB cache.
        UC: UltraChat, OA: OpenAssistant, DD: Dolly Dataset, AD: Alpaca Data.
    }
	\label{fig:multi-model-bench}
\end{figure*}

\textbf{Impact of Memory Pressure:}
Figure~\ref{fig:diff-mem} shows the TTFT of different models under varying background memory pressure.
When the background load is 0GB, both CMA and {\mem} allocations are fast, so {\sys} achieves a similar cold-start TTFT to Strawman-OPT.
However, as memory pressure increases, {\sys} maintains a low TTFT while the TTFT of both Strawman and Strawman-OPT grows rapidly.
This is because high memory pressure leads to excessive page fragmentation, forcing CMA to merge fragmented pages.
Overall, {\sys} achieves an average \textbf{1.87$\times$} speedup over Strawman-OPT, with a maximum of up to \textbf{3.05$\times$}, across memory conditions from 0GB to 10GB.

\subsection{Multi-Model Performance with Cache}

To answer question-3, we evaluate {\sys}'s performance on different model groups and agent workflows, with caching enabled for both model weights and KV caches.

\textbf{Prefill Time}:
We use real-world benchmarks to evaluate the TTFT of four model groups, as shown in Figure~\ref{fig:multi-model-bench}.
Requests are dispatched randomly to models within each group.
{\sys} is configured with a fixed 4GB cache size.
Overall, {\sys} achieves an average \textbf{8.84$\times$} speedup over the Strawman and a \textbf{2.53$\times$} maximum speedup over Strawman-OPT.

\begin{figure}[tb]
    \begin{minipage}[t]{\linewidth}
		\vspace{-0.25cm}

        \centering
        \includegraphics[width=\linewidth]{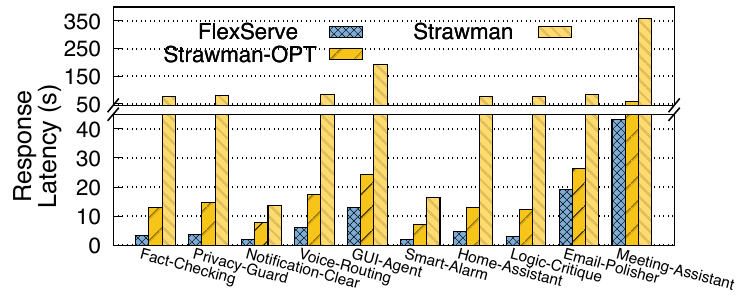}
		\vspace{-0.8cm}

        \caption{Response latency of real-world agent workflows. 
        }
        \label{fig:workflow}
    \end{minipage}
\end{figure}

\textbf{Agent Workflow}:
We evaluate the response latency of ten representative agent workflows on mobile devices, as shown in Figure~\ref{fig:workflow}.
Each workflow may employ a different group of models.
The response latency is calculated as the sum of the last model's TTFT and all previous models' full generation time.
{\sys} improves the response latency of agent workflows by up to \textbf{24.30$\times$} and \textbf{4.05$\times$} compared to the Strawman and Strawman-OPT, respectively.
Some workflows involve a long decode stage, which partially hides {\sys}'s speedup.
Nevertheless, {\sys} still outperforms the Strawman by \textbf{14.15$\times$} and Strawman-OPT by \textbf{2.94$\times$} on average.





\subsection{Overhead to Normal-World Applications}
\label{subsec:eval-normal}

\begin{figure*}[t]
    \centering
    \begin{tikzpicture}
        \node[anchor=south west,inner sep=0] (main) at (-3,0) {
            \includegraphics[width=1.0\textwidth]{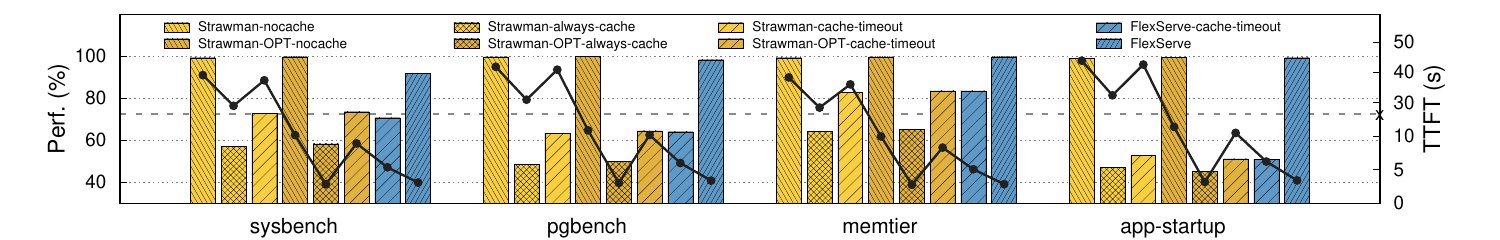}
        };
    \end{tikzpicture}
    \caption{
        Normalized performance of normal-world applications (bars, higher is better) and prefill time of secure inference (lines, lower is better).
    }
    \label{fig:eval-ree-cache}
\end{figure*}




To answer question-4, we evaluate how {\sys} affects the performance of normal-world applications.

\textbf{Influence of Memory Allocation}:
We run SQLite~\cite{sqlite} as the normal-world application, continuously accessing a 3GB in-memory database.
We use Strawman-OPT as the baseline to evaluate how different secure-memory allocation mechanisms affect the performance of REE applications.
Both the Strawman-OPT and {\sys} allocate 8GB of secure memory (or {\mem}) for inference.
During the allocation, Strawman-OPT increases the SQLite latency to \textbf{1.46$\times$} that of {\sys} on average, with a peak of \textbf{2.49$\times$}.
This is because CMA allocation may migrate pages to consolidate free pages into a contiguous block, harming the performance of normal-world applications.

\textbf{Influence of Virtualization}:
We use SPEC CPU 2017~\cite{SPECCPU2017} to evaluate the virtualization overhead imposed on normal-world applications.
The average virtualization overhead is \textbf{2.46\%}.
Moreover, the on-demand protection eliminates this overhead entirely when no secure inference task is active.

\textbf{End-to-End Application Overhead}:
To demonstrate the benefits of {\sys}'s cooperative memory management and LLM-aware memory reclamation, we evaluate the end-to-end performance of both normal-world applications and the secure inference.
A client periodically sends secure inference requests to the secure world, while normal-world performance is measured with four commonly-used benchmarks (\texttt{sysbench}~\cite{sysbench}, \texttt{pgbench}~\cite{pgbench,postgresql}, \texttt{memtier}~\cite{memtier_benchmark}, and \texttt{app-startup}).
Memory is sufficient for either the secure inference or the normal-world benchmarks, and the two are not running concurrently.


Three policies are compared:
1) \emph{no-cache}: release all memory after inference;
2) \emph{always-cache}: retain all model weights/KV caches;
and 3) \emph{cache-timeout}: retain model weights/KV caches for a fixed timeout after inference.
{\sys} uses cooperative memory management to schedule memory between the normal world and the {\sys} Framework.


Figure~\ref{fig:eval-ree-cache} shows the normalized performance of the normal-world applications (bars) and the TTFT of the secure inference (lines).
The normal-world performance is normalized to the baseline without LLM inference.
For the Strawman and Strawman-OPT, the \emph{no-cache} policy reduces the overhead to normal-world applications but increases the TTFT.
The \emph{always-cache} policy achieves a low TTFT but degrades normal-world performance.
The \emph{cache-timeout} policy also fails, since neither the normal nor the secure world is aware of the other's workload and cannot manage memory cooperatively.
In contrast, {\sys} preserves 97.2\% of the non-inference baseline performance for normal-world applications while achieving the lowest TTFT.
It thus attains the benefits of both the \emph{no-cache} and \emph{always-cache} policies simultaneously.

\subsection{Ablation of TTFT Improvement}

\begin{figure}[tb]
    \begin{minipage}[t]{\linewidth}
		\vspace{-0.25cm}

        \centering
        \includegraphics[width=\linewidth]{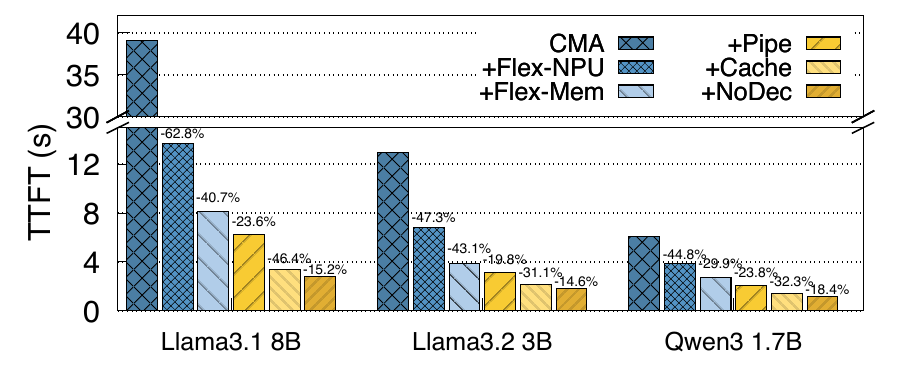}
		\vspace{-0.8cm}

        \setlength{\belowcaptionskip}{-5pt} 
        \caption{
			Ablation of prefill optimization.
        }
        \label{fig:eval-ablation}
    \end{minipage}
\end{figure}

To answer question-5, we evaluate the contribution of individual optimizations to the TTFT.
We measure the TTFT of Llama3.1 8B with 128-token prompt, and enabling different optimizations incrementally.
Figure~\ref{fig:eval-ablation} shows the results.
Each bar reports the TTFT after enabling one additional optimization, and the percentage indicates the incremental TTFT reduction introduced by that optimization.

{\npu} and {\mem} contribute the largest performance gains, reducing TTFT by 49.01\% and 37.89\% on average, respectively.
Even after {\npu}, {\mem}, and pipeline optimizations are enabled, {\sys}'s cache optimization further reduces TTFT by 36.59\% on average.
This confirms the importance of {\sys}'s memory management mechanism.
Although our evaluation platform does not support inline cryptographic hardware, we emulate its performance gain.
It could provide an additional 16.06\% reduction on average.

\section{Discussions}
\label{sec:discuss}

\subsection{Security Analysis}
\label{subsec:discuss-security}

\textbf{Direct Attacks}:
{\sys} considers a kernel-level attacker who tries to compromise the privacy and integrity of the secure LLM inference system.
The code and data of the inference framework reside in either TrustZone's secure memory or {\mem}.
A compromised kernel cannot access them.
{\mem} is also unmapped from the normal-world SMMU, so it cannot be accessed via malicious DMA.
An attacker may attempt to exploit the {\npu} to copy data from {\mem} to unprotected memory.
{\sys} mitigates this by
1) unmapping the {\npu}'s MMIO region from the normal-world kernel; and 2) restricting the {\npu} to access only {\mem} regions.



\textbf{Bypassing {\sys}}:
The protection of {\sys} is based on the S2PT and SMMU, which cannot be tampered or bypassed by the compromised kernel.
For the on-demand protection, the EL3 part of the Flex-Monitor verifies the integrity of the Flex-Monitor's code and data, so attackers cannot modify the Flex-Monitor's code, data, or page tables while the protection is disabled.

\textbf{Response Tampering}:
{\sys} guarantees that responses are generated from the specific user inputs and model weights.
The response proof contains the request hash, the response hash, and a signature signed with the private key of the {\sys} Framework, so any party (e.g., a remote device) can verify the response's integrity.


\textbf{Security Limitations}:
{\sys} shares common security limitations with most TrustZone-based systems.
First, {\sys} cannot protect the normal-world client that sends the inference requests.
If the client is a normal-world application, a compromised kernel can steal or tamper with its input/output, and can even modify the application's code to skip the response-proof verification.
Even so, {\sys} still protects the model weights and KV caches, and a remote device can still verify the response's integrity.
When a secure-world TA sends the inference requests, both the input and the output are protected.
Second, like most TEE systems, {\sys} does not defend against physical attacks, side-channel attacks, or DoS attacks; existing defenses against these attacks are orthogonal to our work and can be applied.

\subsection{Comparison with Different Choices}
\label{subsec:diss-comparsion}

We compare {\sys} with two alternative design choices.
\textbf{Only Using TrustZone}:
A pure TrustZone design must allocate a large, physically contiguous secure memory region, which requires merging fragmented pages and is therefore slow; {\sys} eliminates this merging overhead.
Moreover, the page-granular {\mem} makes it easier for {\sys} to implement cooperative memory management.

\textbf{Using Protected Virtual Machines (pVMs)}:
The inference can also be protected in a protected VM, e.g., the Android pVM~\cite{android-pvm}.
However, such a design incurs additional resource and performance overhead.
First, it must run an OS kernel and a software stack inside the VM, which consumes extra memory; {\sys} instead reuses the existing software stack in the TrustZone secure world.
Second, the VM-based design incurs virtualization overhead even when no secure inference task is running, whereas {\sys} leverages TrustZone's EL3 to implement on-demand protection that mitigates this overhead.

\subsection{Compatibility}
\label{subsec:diss-opt}



\textbf{Compatibility with Other LLM Models}:
Our current implementation supports the Llama3 and Qwen3 series models, and the design of {\sys} is compatible with different model architectures, e.g., the Mixture of Experts (MoE).
New caching optimizations can be added for new architectures, e.g., expert-aware memory management.



\textbf{Compatibility with Other XPUs}:
Currently, {\sys} uses the CPU and NPU to execute inference tasks, but its design also works for GPUs and NPUs from different vendors.
Notably, the NPU used in our evaluation is slower than SOTA NPUs; with a faster NPU, the computation stage would no longer be the bottleneck and the speedup of {\sys} would be even larger.


\section{Related Work}
\label{sec:relate}




\textbf{Secure Model Inference}:
Many existing works~\cite{grover2018privado,ohrimenko2016oblivious,lee2019occlumency,yang2024penetralium,islam2023confidential,jian2025smartzone,li2024translinkguard,mo2020darknetz,shen2022soter,schlogl2020ennclave,xiang2021aegisdnn,elgamal2020serdab,sun2023shadownet} protect model inference from an untrusted OS using a TEE, e.g., ARM TrustZone~\cite{alves2004trustzone} or Intel SGX~\cite{costan2016intel}.
Oblivious ML~\cite{ohrimenko2016oblivious} runs DNN inference in SGX enclaves and uses obfuscation to prevent side-channel attacks.
Confidential DL~\cite{islam2023confidential} protects deep learning inference within TrustZone.
SmartZone~\cite{jian2025smartzone} can also run LLM inference in TrustZone's secure world with multi-threading support.
Other works partition the inference task and run only part of it inside the TEE to achieve higher performance~\cite{li2024translinkguard,mo2020darknetz,shen2022soter,schlogl2020ennclave,xiang2021aegisdnn,elgamal2020serdab,sun2023shadownet}.
TransLinkGuard~\cite{li2024translinkguard} runs the locked model outside the TEE and protects only an authorization model in the TEE.
Darknetz~\cite{mo2020darknetz} runs only part of the DNN layers in TrustZone's secure world to balance security and performance.
These works do not address the challenges of inflexible isolation and inefficient resource management in TrustZone, discussed in Section~\ref{subsec:motiv-challenges}; thus, they cannot achieve both high LLM inference performance and low normal-world application overhead.


TZ-LLM~\cite{wang2025tz} protects LLM inference with TrustZone and introduces a pipelining method to hide the latency of allocating contiguous secure memory.
Rather than hiding the merging overhead of CMA allocation, {\sys} introduces page-granular {\mem} to eliminate the merging procedure entirely.
{\sys} further introduces cooperative memory management, which allows the normal-world OS and the secure-world inference framework to jointly manage memory efficiently.
TZ-LLM does not address the challenge of inefficient resource management.
As a result, {\sys} achieves both high secure LLM inference performance and low normal-world application overhead.

ASGARD~\cite{moon2025asgard} protects on-device DNNs inside a protected VM, which is constructed based on Linux pKVM~\cite{linux-pkvm} and Android pVM~\cite{android-pvm}.
As discussed in Section~\ref{subsec:diss-comparsion}, it introduces additional resource overhead for the guest VM, and cannot reuse existing software stack of the TrustZone.
The NPU is passed through to the protected VM.
Meanwhile, ASGARD does not address the challenge of secure memory management in {\sys}.

\textbf{Enabling Accelerators in TEE}:
Existing works also try to enable accelerators, e.g., GPUs and NPUs, inside the TEE~\cite{fan2025xputee,volos2018graviton,jang2019heterogeneous,deng2022strongbox,zhu2020enabling,wu2023building,park2023safe,mai2023honeycomb,sageATC,Telekine}.
Graviton~\cite{volos2018graviton} modifies the GPU chip to support a GPU TEE.
StrongBox~\cite{deng2022strongbox} targets ARM platforms with integrated GPUs without requiring hardware modifications.
SAGE~\cite{sageATC} ensures a verification function is securely deployed on the GPU, though a compromised OS can still access the GPU arbitrarily.
XpuTEE~\cite{fan2025xputee} provides a high-performance heterogeneous TEE for high-performance GPUs.
Unlike existing works, which primarily focus on isolating accelerators, {\sys} addresses the problem of efficiently switching between the unprotected NPU and {\npu}.


\section{Conclusion}

This paper presents {\sys}, a fast and secure device-side LLM inference system that defends against an untrusted OS kernel.
{\sys} first constructs {\mem} and {\npu}, which can only be accessed by the secure world yet can be efficiently allocated and reclaimed by the normal-world OS.
Furthermore, the {\sys} Framework runs secure LLM inference in the secure world based on {\mem} and {\npu}, and performs cooperative secure memory management with the normal-world OS.
We compare {\sys} with two TrustZone-based Strawman designs.
The results show that {\sys} achieves average TTFT speedups of 10.05$\times$ over the strawman and 2.44$\times$ over an optimized strawman.



\bibliographystyle{ACM-Reference-Format}
\bibliography{refs}

\end{document}